\documentclass[twoside]{dis09}
\usepackage[latin1]{inputenc}
\usepackage[dvips]{graphicx,epsfig,color}
\usepackage{wrapfig,rotating}
\usepackage{amssymb,amsmath,array}

\begin{document}
\title{Future Facilities Summary}

\author{Albert De Roeck$^1$ and Rolf Ent$^2$
%
%
\vspace{.3cm}\\
%
1- CERN  \\
1211 Geneva 23 Switzerland\\
and Universiteit Antwerpen, Belgium \\
%
\vspace{.1cm}\\
2- TJNAF \\
Newport News, Virginia 23606, USA\\
}

\maketitle

\begin{abstract}
For the session on future facilities at DIS09 discussions were organized  on  
DIS related measurements that can be
expected in the near and medium --or perhaps far--  future, including plans
from JLab, CERN and FNAL fixed target experiments, possible measurements and detector upgrades at RHIC,
as well as the plans for  possible future electron proton/ion colliders such as the EIC and the LHeC project.

\end{abstract}

\section{Introduction}
The field of Deep Inelastic Scattering (DIS) and related QCD topics is an extremely active one, as demonstrated
by the program at this years'  DIS meeting. This report gives a brief summary of the session on the 
possible future 
experimental program. 
The very lively discussions during the sessions are a good testimony of the continuing strong interest 
of the proponents in this field for a next generation of data, experiments and facilities.
During the final plenary day of the meeting 
a panel discussion was organized on two possible future electron-proton/ion collider 
projects, namely the Electron Ion Collider (EIC), as proposed for a possible new facility  in the US, with 
interested laboratories BNL and JLab, and the Large Hadron-electron Collider (LHeC), a study encouraged 
by ECFA in Europe.

In detail the program of the meeting was
\begin{itemize}
\item
JLab session (\cite{ent,liyanage, guidal, keppel, liuti}):
Towards the 12 GeV electron beam upgrade and the 
experimental programs in Halls A/B/C and D
\item
Compass: High energy muon-scattering at CERN  (\cite{magnon})
\item
Minerva:  Neutrino scattering at FNAL (\cite{ziemer})
\item
E906: Drell-Yan at FNAL (\cite{reimer})
\item
RHIC upgrades (\cite{nouicer,fukao,lee}), which include
muon triggers for W tagging, proton tagging
heavy flavor tagging
\item
The Electron Ion Collider EIC project\cite{caldwell,lamont,kinney,weiss,livitenko,aschenauer}
\item
The Large Hadron electron Collider LHeC project\cite{stasto,behnke,klein,rojo,holzer,pollini}
\end{itemize}

\begin{figure}[t]
\centerline{\includegraphics[width=0.45\columnwidth]{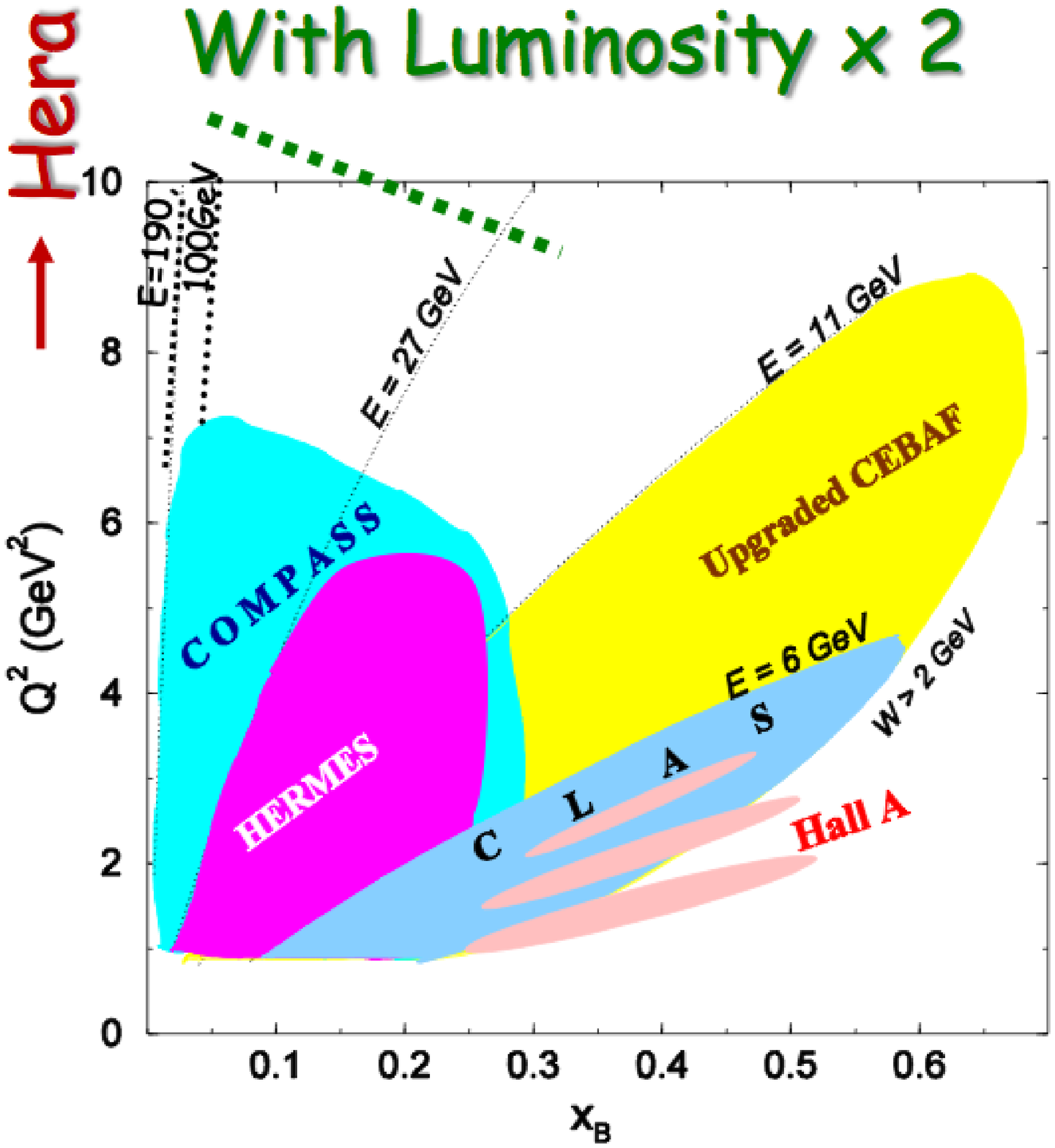}\includegraphics[width=0.45\columnwidth]{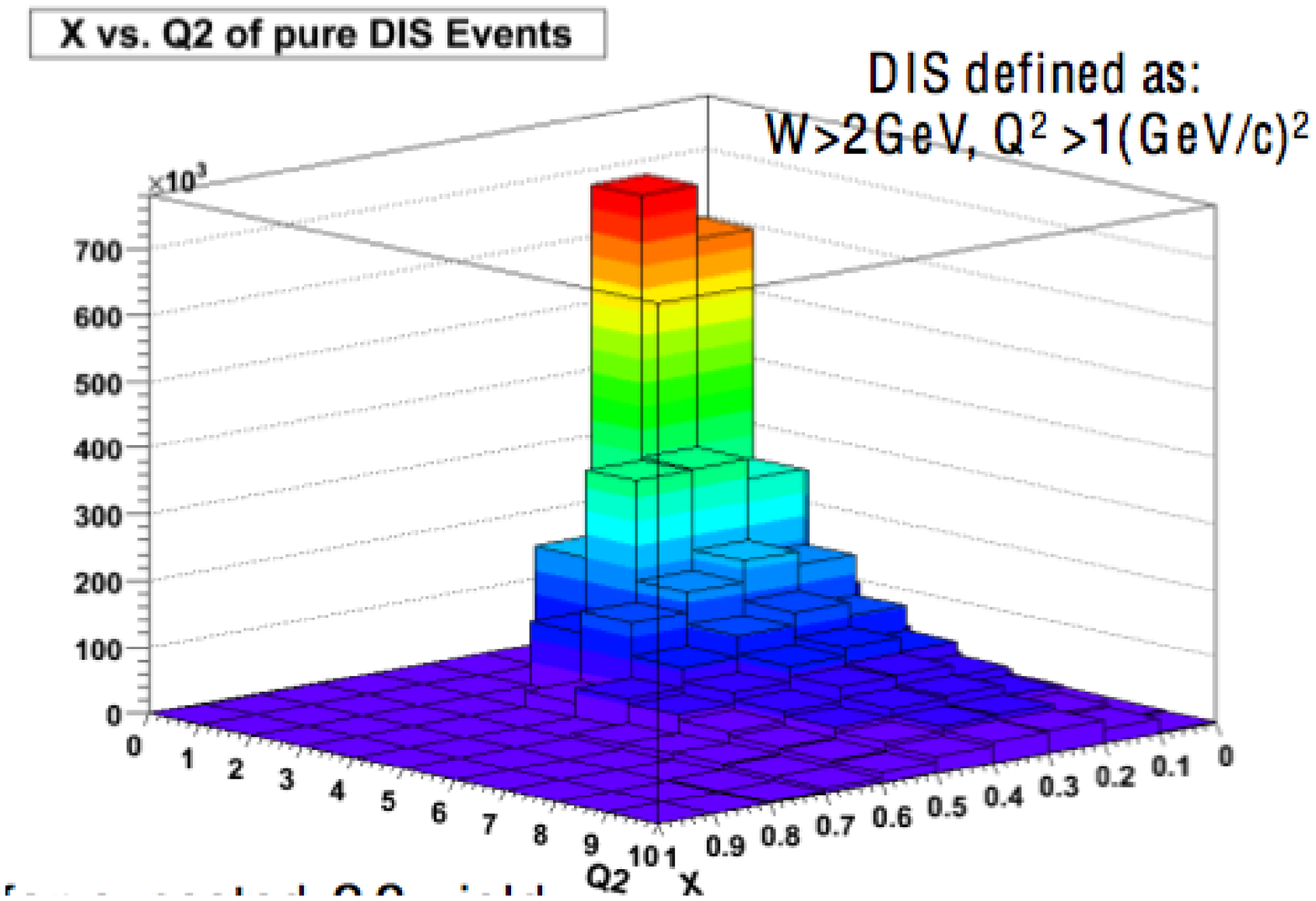}}
\caption{(a, left) The kinematic range of DIS fixed target experiments, including the reach of COMPASS with
increased luminosity. (b, right) The expected kinematic coverage for MINERvA DIS events}\label{Fig:compas}
\end{figure}

\section{Fixed Target Experiments}
A survey was made on the near future fixed target experiments program at JLab, FNAL, and CERN.

At JLab the construction phase of the 12 GeV Upgrade has started, and the commissioning is expected to 
start in 2014. The upgrade will also include a new hall D and upgrades in the existing halls A, B, and C.
An overview of the hall capabilities is shown in Table~\ref{jlab}.

\begin{table} [h]
\begin{center}
\begin{tabular} {|c|c|c|c|} \hline

Hall D &	Hall B &	Hall C & Hall A  \\ \hline
excellent hermeticity & 	luminosity $10 . 10^{34}$	 & energy reach  & instalation space \\ \hline 	
polarized photons	& hermeticity	& precision & \\ \hline  	
Eg~8.5-9 GeV	& \multicolumn{3}{|c|} {11 GeV beamline} \\ \hline
$10^8$ photons/s	& \multicolumn{3}{|c|} {target flexibility} \\ \hline
\multicolumn{2}{|c|} {good momentum/angle resolution} &	\multicolumn{2}{|c|} {excellent momentum resolution} \\ \hline 
\multicolumn{2}{|c|} {high multiplicity reconstruction} & \multicolumn{2}{|c|} {luminosity up to $10^{38}$} \\ \hline	
\multicolumn{4}{|c|} {particle ID} \\
\hline

\end{tabular}
\caption{Overview of the hall capabilities at JLab.}
\label{jlab}
\end{center}
\end{table}

The physics program for the different experimental halls is as follows 
\begin{itemize}
\item
Hall D: exploring the origin of confinement by studying exotic mesons.
\item
Hall A:  spin structure, form factors, future new experiments. \\
Building on surprising results in elastic scattering, strangeness form factors, spin asymmetries at large $x$ \\
$\rightarrow$ Include precise measurements of axial quark couplings, precision low-energy Standard Model tests at the ILC mass scale (through Parity-Violations).
\item
Hall B:  understanding nucleon structure via generalized parton distributions.\\
Building on merging information of exclusive and inclusive measurements \\
$\rightarrow$ program on Generalized Parton Distributions (Deep-Virtual Compton Scattering!), 
transversity studies, d/u at large $x$, hadron holography.
\item
Hall C: precision determination of valence quark properties in nucleons and nuclei \\
Building on studies of quark-hadron duality, structure functions at large $x$, and precision L/T separations \\
$\rightarrow$  L/T separations in semi-inclusive and deep exclusive charged-meson production, factorization tests, flavor $k_T$-dependence in SIDIS, hunt for  so called superfast quarks.
 \end{itemize}

 COMPASS is a facility at the CERN SPS that will continue to study DIS with muon beams of 100 GeV to 200 GeV on  fixed targets. An LOI was submitted on future measurements related to the nucleon spin structure.
 Some topics need detector upgrades and perhaps beam intensity upgrades.
Kinematically COMPASS is nicely complementary 
to the 12 GeV beam reach at JLAB as shown in Fig.~\ref{Fig:compas}(a) 
 for GPDs and transverse spin studies. The reach extends  to low-$x$. 
 Possible measurements for COMPASS include  
 \begin{itemize}
 \item
 Precision measurements of the transverse spin effects in semi-inclusive DIS
 \item
 Measuring the spin dependent longitudinal structure function at small $x$
 \item
 Study of Generalized Parton Distributions
 \item
 Study of transversity distributions in Drell-Yan
 \end{itemize}
 For the first two subjects modest detector upgrades are foreseen, while for the last two important hardware 
 upgrades are envisaged.
  


 \begin{figure}[t]
\centerline{\includegraphics[width=0.5\columnwidth]{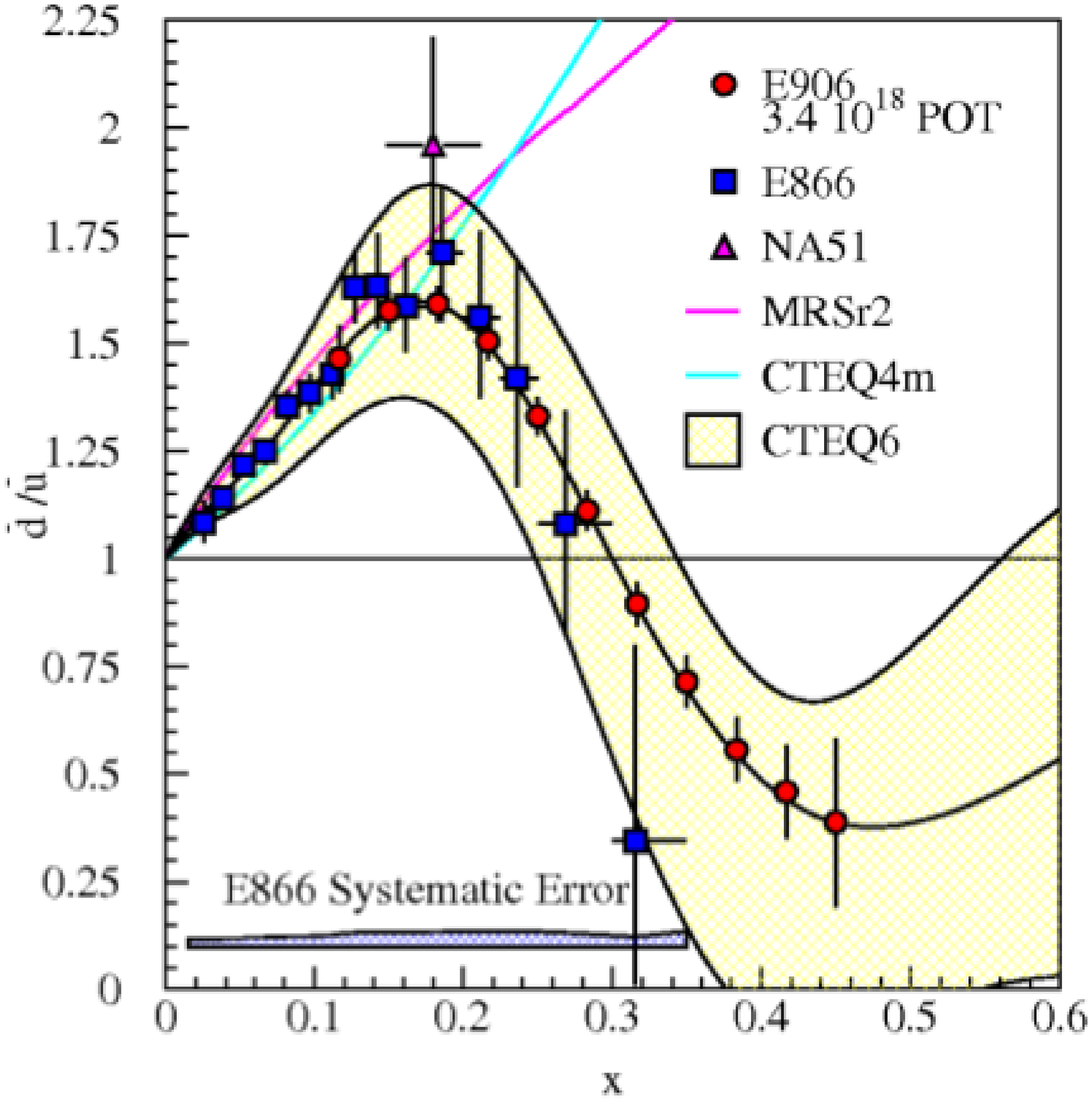}\includegraphics[width=0.5\columnwidth]{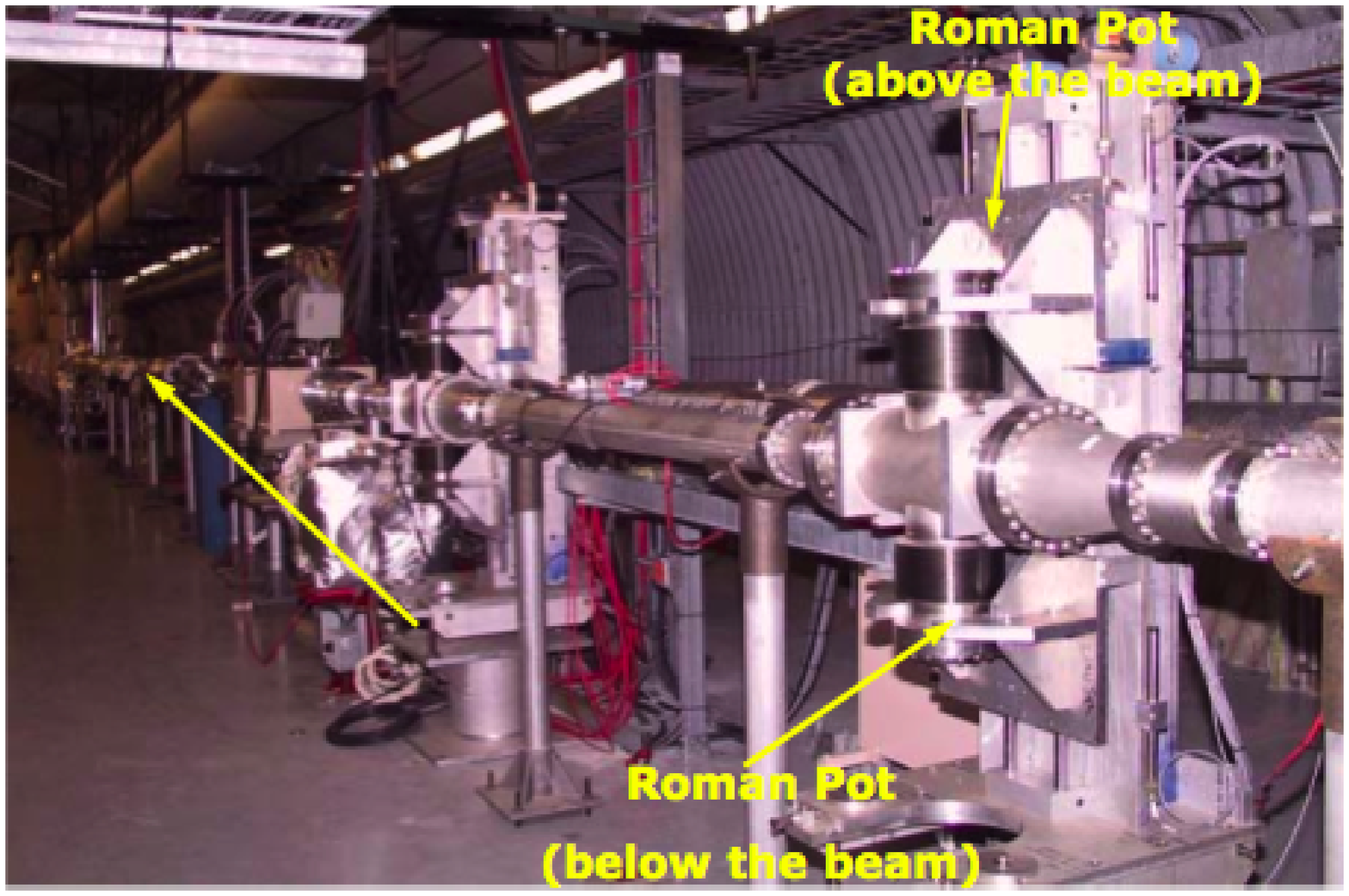}}
\caption{(a,left) The expected kinematic range and precision for the E906 experiment. (b, right)
The roman pots from the pp2pp experiment now moving to STAR. }\label{Fig:fnal}
\end{figure}

 The MINERvA experiment will use neutrino beams on nuclear targets at FNAL and is set to make 
 precise measurements of neutrino cross sections with neutrino beams of energies up to roughly 30 GeV.
 The DIS coverage is shown in Fig.~\ref{Fig:compas}(b).
 MINERvA will collect 6M events on a carbon target in the transition (not so deep DIS) and DIS region plus an additional  6.5 M events in four nuclear targets. Different specific studies will focus on various regions of 
 the kinematical variable space,  and MINERvA 
 will significantly increase the existing neutrino data set available to the community.
 For DIS MINERvA will measure
 \begin{itemize}
 \item
 Cross sections in the transition region to test quark-hadron duality
 \item
 Structure function ratios for combinations of nuclear targets
 \item FSI via nuclear target multiplicities and hadron shower energies
 \end{itemize}

 The E906 collaboration at FNAL is set to measure Drell-Yan production via muons. Drell-Yan scattering is
 uniquely sensitive to the antiquark distributions of the target. E906 will measure 
 \begin{itemize} 
 \item The ratio of the $\overline{d}$  to $\overline{u}$ distributions in the proton
 \item The modifications to the quark sea in a nucleus
 \item and much more...
 \end{itemize}
 The expected statistics that will be collected is a factor 50 larger than that of E866/NuSea. 
 The incoming proton beam 
 energy will be only 120 GeV, reducing in the energy squared $s$ by a factor of 7 with respect to 
 E866/NuSea. Thus
 the measurements of E906 will cover a different kinematic range, namely
 up to  high $x$  values of 0.5, as shown in Fig.~\ref{Fig:fnal}(a).
 The run of E906 is scheduled to start in 2010 and to last for 2 years.
 
\section{Hadron Colliders: RHIC}
The hadron collider RHIC is entering a new phase. Initial measurements of the spin program have been
performed, and have  led to a measurement of the polarized gluon distribution 
$\Delta G$ in a restricted range of Bjorken- $x$. The next phase will 
include improvements in precision, an extended kinematical range,  and new measurements, such as those of heavy flavor production.
In order for these measurements to be made, PHENIX plans to add a Silicon Vertex Detector in the barrel
and forward region. The most inner layer of the barrel will be at 2.5 cm away from the beam.
A new technology, called silicon stripixel will be used for this detector. The installation for the barrel is expected to be completed in the fall of 2010, and for the forward detectors in fall of  2011.

PHENIX also plans a forward $W$ trigger  upgrade  (RPC + MuTRG) 
to study the sea quark polarization in proton. With the polarized beams at RHIC
the $W$ production gives access to the polarized
valence quarks and  anti-quark distributions in the proton. The statistics is however an important issue for
such an analysis. The new trigger should allow for an additional rejection factor of 100 to purify
the muon sample at the trigger level, allowing for a higher rate to be accepted.
The development of the new RPC and MuTRG is ongoing. A part 
of the detectors were already installed during shutdown period in 
2008.  RPC and MuTRG data were taken and used to  evaluate the performance of  the detector, which 
was found to be very satisfactory. The installation of the RPC and MuTRG will be completed by the 
next physics run at 500 GeV.

STAR reported on a planned future physics program with tagged forward protons. The detectors are those
used before in the pp2pp experiment in 2002-2003, and are shown on in Fig.~\ref{Fig:fnal}(b). 
 In phase-I 8 roman pots will be installed at 55.5-58.5 m from the interaction point. 
These should be ready in 2009 for the data taking run. In a phase-II of the project
8 (12) roman pots are planned to be installed in 2010-2011 at 13/16m. During phase-I the acceptance 
in $|t|$ range will be 0.002- 0.2 GeV$^2$ and for phase-II the acceptance will be extended 
up to 1.3 GeV$^2$.
 The addition of these detectors will 
clearly enhance the physics capabilities of STAR for eg peripheral collisions and for central exclusive
production processes.

\section{Future Electron proton/ion Colliders}

Interest is growing worldwide in a novel electron-ion collider. Design concepts exist at CERN, with ideas to intersect an electron accelerator with the Large Hadron Collider (LHeC), and in the U.S. to either add an electron accelerator to the Relativistic Heavy Ion Collider  at BNL, or an ion accelerator to the upgraded 12-GeV Continuous Electron Beam Accelerator Facility at JLab. The US project is generically called the EIC.

\begin{figure}[t]
\centerline{\includegraphics[width=0.7\columnwidth]{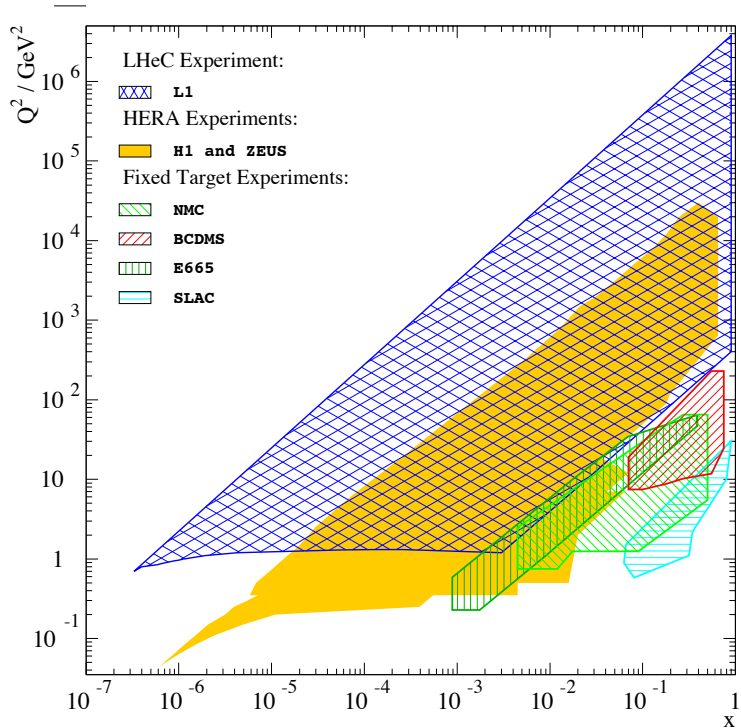}}
\caption{The kinematic reach for LHeC}\label{Fig:LHeC_kinem}
\end{figure}

The future DIS working group devoted 3 subsessions where the LHeC and EIC were discussed in detail.
Possible machine lay-outs and parameters were presented, and the physics program was addressed. 
Clearly the high energy of the LHeC allows to explore a new kinematic area for both ep and e-ion
collisions. The high luminosity anticipated for the EIC and the possibility for polarized beams will allow for number a precision and novel measurements  as discussed in the various contributions in this session.

The LHeC has two alternative scenarios: a ring-ring (RR) scenario and a linac-ring (LR) scenario. For 
the RR scenario one typically  has a 50 GeV electron beam on a 7 TeV proton beam (or 2.75 TeV heavy ion 
beam), and a peak luminosity of around $5 . 10^{33}$ cm$^{-2}$s$^{-1}$ for 50 MW power.
The LR scenario has the  potential to reach larger electron energies, perhaps up to 150 GeV, but in general
the total integrated luminosity will by a factor 5 to 10 lower compared to the RR option.

The EIC projects discussed by BNL and JLab describe an electron beam of 4 to 20 GeV on a proton beam of
50 to 250 GeV. The peak luminosity aimed for is similar to the RR LHeC option. Polarization is an integral part of the proposal, aiming for 70\% of polarization for each beam.

\begin{figure}[t]
\centerline{\includegraphics[width=0.65\columnwidth]{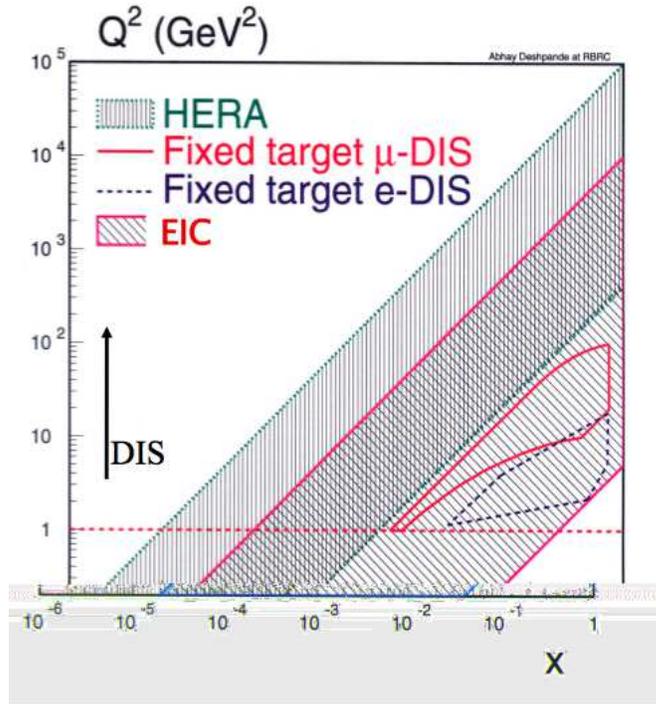}}
\caption{The kinematic reach for  EIC}\label{Fig:EIC_kinem}
\end{figure}

The physics program of both these colliders is discussed in detail in the contributions
\cite{caldwell,lamont,kinney,weiss,stasto,behnke,klein,rojo} during the session.
The kinematic reach covered by the LHeC is shown in Fig.~\ref{Fig:LHeC_kinem} and for the 
EIC in Fig.~\ref{Fig:EIC_kinem}.

In order to probe the interest of the community and introduce the projects to wide audience a plenary panel discussion was organized starting off with an introduction of Albert De Roeck who presented a number of 
question for the discussion such as
\begin{itemize}
\item Is there sufficient backup in the community for such projects?
\item Is there a scientific preference in the community for a specific project?
\item For the projects themselves: what are the possible timelines, support in 'host' labs, and possible next steps?
\item What are possible stage projects and how interesting are these?
\end{itemize}

The panel discussion then proceeded
with presentations on both the Large Hadron Electron Collider (LHeC) by Max Klein\cite{mklein}  and the U.S. Electron-Ion Collider (EIC) by Abhay Deshpande\cite{deshpande}. This was followed by short exposes about their personal views on these efforts by Allen Caldwell (chair of the LHeC scientific advisory committee and member of the EIC advisory committee), Tony Thomas (chief scientist at Jefferson Laboratory) and Steve Vigdor (head of high-energy and nuclear physics division at BNL), followed by a general discussion with questions from the audience. Unfortunately, Sergio Bertolucci, CERN scientific director, could not attend. Outlined below is an (incomplete) description of the discussions following the separate short exposes, where applicable, and the general discussion after all presentations. More detailed descriptions on the introductory overview presentations on the LHeC and EIC status and plans, their science cases and accelerator and detector plans, and the short personal view presentations by panel members are given elsewhere in the proceedings. 

First of all, it was commented that there also exists non-U.S. and non-European interest in the EIC and LHeC ideas. It was questioned whether any tension, conflict or interaction exist between the LHeC and the next Linear Collider ideas, but this will depend on highly-anticipated LHC results. Both U.S. EIC design efforts have considered staging plans, and the question arose how to decide amongst such. This depends truly on their science motivation as any staging must have a standalone case to be successful. An interesting comment from the floor was that the common feature between the LHeC and EIC ideas was really only their high luminosities, as energies, beam polarization, and ion species differ, and furthermore they aim at different science goals. This is certainly true. It was pointed out, however, that they have costs in common: if final designs were found to be expensive it is debatable if one can have two such colliders. In addition, the experimental facilities and techniques would be similar. Certainly, the U.S. EIC ideas and the LHeC will benefit from sharing in items like detector and interaction region ideas, and simulation concepts. Discussed together also gives a chance to better gauge their physics communality, e.g., to access parton distribution functions at low Bjorken x. 
Concerning the costs, it was later mentioned to be wary of requiring a facility that meets all scientific goals, as one can price oneself out of the market. A similar concern was raised about the detector ideas, to be wary of the possibility to develop one universal super detector. It may be more effective in terms of cost, feasibility and even science access to build a few specialized detectors rather than one to "do it all".

A question arose on what the viewpoint of CERN was towards an LHeC. Discussions with the CERN general director indicated the desire for the development of a (very good) Conceptual Design Report. Number one priority within CERN is of course the LHC startup, but large support for an LHeC already exists within the CERN accelerator groups. CERN has invited the LHeC proponents to develop international collaboration, which is ongoing with recently an example of work with Novosibirsk on the design for magnets for both the electron ring and the detector. CERN is still far away from a next big project, but this also allows the opportunity to shape the science future in terms of validating the case of an electron-proton collider in combination with existing p-p and A-A colliders. It is expected that all science opportunities, including an LHeC and an LHC upgrade, should be on the table with CERN management around 2012-2013, such that the future plans can be decided upon. It was noted that there is in addition support for the LHeC science case development by both ECFA and NuPECC, the latter having its long-range planning process next year and study groups dedicated to both LHeC and EIC.

As a next step, the LHeC proponents have a workshop from September 1-3 in Divonne near CERN   to work on a Conceptual Design Report. To support the EIC efforts, several topical workshops are planned beyond the regular EIC Collaboration meetings that occur twice a year, with the next collaboration meeting end of May at GSI/Darmstadt. A one-week long workshop to further discuss and develop the scientific case of a staged EIC is scheduled for October 2009 at the U.S. Institute for Nuclear Theory in Seattle, WA. A two-month program follows in the Fall of 2010. Since the U.S. Department of Energy only endorses a Conceptual Design Report after awarding a "mission need" statement about a new facility, the goals of these workshops are white papers to assist preparations for the next U.S. long-range planning effort, also expected in 2012-2013. A joint timeline for the EIC exists to come up with the best possible EIC design and science case by this next long-range plan. The efforts to look at staged versions of an EIC originated from the hope to possibly have such an EIC, with a standalone and excellent science case to support the costs, around 2020.

In reply to a question on the LHeC, its timeline is beyond 2020: it was stated that it takes two years to further develop a Technical Design Report from a completed Conceptual Design Report, and then about eight years to build a detector. This sounds like a long period, but silicon vertex trackers for DESY experiments were proposed in 1992, with results only shown at this DIS conference. If one does not start development now, an LHeC certainly will not happen on the anticipated time scale. At present, the costs still need to be developed for all LHeC scenarios considered (a range of costs was discussed for the various EIC scenarios) and will need to be included in the Conceptual Design Report, including also installation costs. Whatever design scenario is chosen, the detector will be a huge thing! It was mentioned that questions exist that cannot be answered yet, with associated cost uncertainties, such as how to bypass existing LHC detectors in the LHeC ring-ring option. Similarly, the ring-linac option costs will scale with the linac length or energy. LHC results will give guidance on the proper energy range for an LHeC.

Regarding the EIC, a question was raised on the advantages and disadvantages of the choices for the staged EIC considerations. It was noted that the energy ranges for an eventual high-energy EIC, either at BNL or at JLab are similar. The main difference between the staged EIC options comes from the repetition rates and beam parameters assumed, with further impact on the exact interaction region and detector design. In general, the geometry of the detector/interaction regions is presently assumed to be similar. Detector designs are still at an early stage and further community involvement in developing these in more detail was solicited. For either staged-scenario design, it is not clear at all yet to what extent the detectors can also serve for a future higher-energy EIC. The advantage of a higher-repetition rate would be the higher resulting luminosity.

During the discussion it was emphasized at several occasions that there is plenty of ground for the EIC and 
LHeC to collaborate. In particular the tools for ep and eA scattering, such as state of the art Monte Carlo 
Generator programs, and generic detector design and detector technology evaluation/studies would 
be very suited for common discussions and perhaps even working groups. This would certainly be very resource-efficient and also demonstrate the  unity and determination in the DIS world.

The panel discussion was summarized stating that this is an important time for the worldwide efforts to establish an electron-ion collider. The LHeC efforts gear up to complete a Conceptual Design Report around 2012-2013, whereas the EIC is working on substantiating their science case and accelerator/detector design for the next U.S. Nuclear Science Advisory Committee long-range planning effort in the same time frame. This is the right time to join in to form a critical mass, to develop a sufficiently strong science case, and do detector work and simulations. It is very important to have strong backup of the laboratories, and to develop task forces. A favorable time may very well be coming up for such electron-ion collider ideas.

Finally, the chair John Dainton thanked the laboratory representatives, the panel members, and the audience for the discussions.

\section{Conclusions}
At the DIS09 meeting there was a very clear and strong sign from the community in continuing and growing interest in a DIS program in the future. Several new experiments will come on line in the next years, which will 
provide a wealth of new data and will lead to new insights. The LHeC and EIC projects are shaping up to 
become conceptual designs. The success of these projects will, among others, depend on the continuing commitment and engagement of the DIS community to these projects now and in the next few years. If a window of opportunity would present itself, we better not miss it.

\begin{footnotesize}




\begin{thebibliography}{99}
\bibitem{ent} R. Ent, The JLAB 12 GeV upgrade- Short introduction, these proceedings
\bibitem{liyanage} N. liyanage, The Hall A deep inelastic scattering program at 12 GeV, these proceedings.
\bibitem{guidal} M. Guidal, The CLAS12 detector at JLAB to measure GPDS and large-x PDFs, these proceedings.
\bibitem{keppel} C. Keppel, The SHMS and the Hall C L/T separated DIS program, these proceedings
\bibitem{liuti} S. Liuti, The 12 GeV upgrade DIS related program, these proceedings.
\bibitem{magnon} A. Magnon, Future plans of COMPASS, these proceedings
\bibitem{ziemer} B. Ziemer, The Minerva detector and science possibilities, these proceedings.
\bibitem{reimer} P. Reimer, Exploring the Anti-quark Structure of Matter with Drell-Yan, these proceedings
\bibitem{nouicer} R. Nouicer, Vertex detector upgrades for the Phenix detector, these proceedings.
\bibitem{fukao} Y. Fukao, The Phenix forward muon trigger upgrade project for the study of the proton spin, these proceedings.
\bibitem{lee}J.H. Lee, Future physics program at RHIC with tagged forward protons, these proceedings.
\bibitem{caldwell} A. Caldwell, ep and eA physics at low Q2, these proceedings.
\bibitem{lamont} M. Lamont, eA physics at a future Electron-Ion Collider, these proceedings.
\bibitem{kinney} E. Kinney, Spin structure studies at an EIC, these proceedings.
\bibitem{weiss} C. Weiss, Physics with a low/medium energy EIC, these proceedings
\bibitem{livitenko} V. Livetenko, Designs for an Electron-Ion Collider, these proceedings
\bibitem{aschenauer} E.C. Aschenauer, EIC detector plans, these proceedings.
\bibitem{stasto} A. Stasto, QCD at the LHeC, these proceedings.
\bibitem{behnke} O. Behnke, Precision studies of QCD and EW interactions, these proceedings
\bibitem{klein} U. Klein, New Physics at large scales, these proceedings.
\bibitem{rojo} J. Rojo, Potential of the LHeC to constrain PDFs, these proceedings
\bibitem{holzer} B. Holzer, LHeC facility plans, these proceedings.
\bibitem{pollini} A. Pollini, LHeC detector plans, these proceedings
\bibitem{mklein} M. Klein, The LHeC Project, these proceedings.
\bibitem{deshpande} A. Deshpande, The EIC Project, these proceedings.

\end{thebibliography}
%

\end{footnotesize}

\end{document}